# Novel mesophase behavior in two-dimensional binary solid solutions

B.P. Prajwal and Fernando A. Escobedo*

Department of Chemical and Biomolecular Engineering, Cornell University, Ithaca, New York 14853, USA

**Abstract**

Monte Carlo simulations were used to study the assembly of binary mixtures of hard disks with squares, where the components have size ratios that optimize their co-assembly into compositionally disordered solids. It is observed that, along with the enhanced regions of solid miscibility, a continuous-looking transition from the disk-like to the square-like behavior occurred through a novel mosaic (M) phase, which seamlessly bridges the regions of hexatic mesophase of disks and the tetratic mesophase of squares. The M phase has interspersed tetratic, hexatic, and rhombic-like locally ordered clusters.

Recent advances in the synthesis [1–3] and fabrication [4,5] of faceted sub-micron particles with different shapes have spurred interest in their use as building blocks for the assembly of targeted complex structures. Several tunable parameters like particle shape [6,7] and inter-particle interactions [7,8], allow the design of a wide range of morphologies having enhanced optical characteristics for potential applications in nanophotonics [9,10], sensing [11], and catalysis [12–14]. Towards designing such materials, recent efforts have focused on predicting phase behavior using theory [15,16] and simulation [6,17–21] for hard polyhedral particles in the bulk (3D) and in monolayers (2D), where the formation of ordered structures entirely depends on the entropic forces encoded in the particle shape. In particular, several experimental protocols leveraging slit confinement or interfacial pinning [22,23] can be deployed to assemble monolayers from different readily synthesizable nano- and micro-sized polyhedral or polygonal particles for applications in thin-film optical and electronic devices [24–28].

Single-component hard-particle superstructures arise at sufficiently high concentrations due to packing entropy manifesting as effective entropic bonds between the constituent particles. Pure systems of squares have been predicted to exhibit a Kosterlitz-Thouless-Halperin-Nelson-Young (KTHNY) behavior, wherein the transition is continuous between both isotropic fluid and tetratic phase and tetratic and solid phases [21]. Simulation results reported for the melting behavior of hard disks suggest that the transition occurs in two steps with a first-order fluid-hexatic transition and a continuous hexatic-solid phase transition [29]. The tetratic and hexatic phases are partially ordered mesophases characterized by a short-range translational order and quasi-long/long-range bond orientational order.

By *'mixing'* particles of different shapes, we can access a wider variety of superstructures having a combination of the constituents' physical properties. For example, ordered superstructures have been predicted for binary mixtures of hexagons+squares, squares+triangles, hexagons+triangles with and without enthalpic patchiness encoded in their facets [30]. The phase behavior of binary mixtures strongly depends on the relative size ratios and contents of the components. This correlation was observed in a size-bidisperse system of hard disks, where the liquid-hexatic-solid transition changes to a first order liquid-solid transition upon increasing the composition of the small disks [31]. For binary mixtures of parallel hard squares having disparate sizes, a fluid-solid phase-separated state was found with small and large squares forming the fluid and solid phases, respectively [32]. These predicted phases reflect the interplay of mixing and packing entropy. At very high pressures, packing entropy dominates over mixing entropy leading to strong segregation of the components into their respective stable structures.

The focus of this paper is to explore the phase behavior of 2D hard binary mixtures of disks+squares, when the components have size ratios that optimize their co-assembly into solid solutions. The size ratio is defined as $\xi = \sigma/a$ where $\sigma$ = disk diameter and $a$ = square edge length. For this purpose, we adopted the exchange free-energy method [33] to predict $\xi$ values which tend to maximize the range of compositions and packing fractions where substitutionally disordered solid solutions occur. This general approach was recently introduced and applied to 3D mixtures of spheres and polyhedra. The method is based on finding the $\xi$ value that minimizes an exchange free-energy ($\Delta F_x$) metric, which is obtained by adding the excess chemical potentials associated with substituting one particle in each pure host solid by a guest particle:

$$\Delta F_x = \mu_{ex}^{S_1}(host \to guest) + \mu_{ex}^{S_2}(host \to guest) \qquad (1)$$

where $\mu_{ex}^{S_i}$ is the reduced excess chemical potential (in units of thermal energy) associated with a single-particle host-to-guest mutation in pure phase $s_i$ ($i = 1$ or 2). $\mu_{ex}^{S_i}$ is also a mixing free energy at infinite guest dilution (see connection in Supplementary Information, SI, Sec. I) and hence by minimizing $\Delta F_x$ i.e., the *"cost"* for host-guest substitutions in both solid phases, mixing entropy and substitutionally disordered solution behavior are enhanced.

The disk+square mixture with optimized $\xi$ was found to exhibit a novel mosaic (M) phase having locally ordered microscopic clusters with square-rich four-fold and rhombic (RB) lattice symmetry, and disk-rich six-fold symmetry, that are distributed randomly throughout the simulated domain. This unique behavior of coexisting finite clusters of two different symmetries can be seen as a mesophase bridging the hexatic and tetratic mesophases observed for the disk-rich and square-rich systems, respectively.

We verified the formation of solid solutions by mapping the pressure-composition phase diagram using hard-particle Monte Carlo simulations in the isothermal-isobaric ensemble (see SI Sec. II) for the *optimized* components size ratio $\xi= 1.1$ (see SI Sec. III). The phase boundaries were identified by analyzing the local correlation of the six-fold and four-fold bond-orientational (see SI Sec. IV for details) and the positional order parameters. At high pressures, the mixtures phase separate into their respective nearly pure component solid phases. The regions where the two phases coexist were mapped based on the results from interfacial simulations (see SI Sec. II). Most interfacial simulations were carried out at the equimolar global composition, with additional runs performed for other compositions to better map out the two-phase coexistence boundaries. Results are reported in dimensionless quantities for distance, $r^* = r/a$, reduced pressure, $P^* = Pa^2/k_bT$ and area fraction/density, $\eta = NA_p/A$, where $P$ is pressure, $k_b$ is Boltzmann's constant, $T$ is temperature, $N$ is the total number for particles, $A$ is the total area of the system, and $A_p$ is the area occupied by the particles.

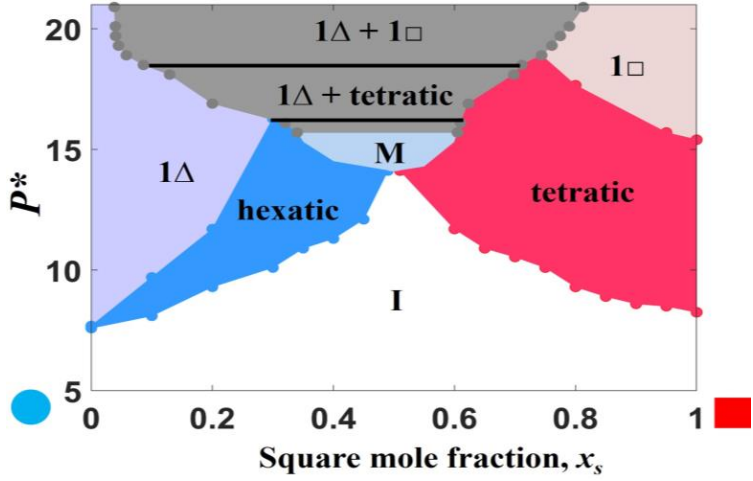

**FIG. 1** (color online). Pressure-composition phase diagrams for mixture of disks+squares with *optimal* component size ratio, $\xi = 1.1$. The symbols $1\Delta$, $1\square$, I, and M denote the triangular solid, square solid, isotropic and the mosaic phase, respectively.

The pressure-composition (Fig. 1) and area fraction-composition (Fig. S3) phase diagrams exhibit broad stable regions of substitutionally disordered square-rich $1\square$ (square lattice) and disk-rich $1\Delta$ (triangular lattice) solid solutions along with the hexatic (in disk-rich region) and tetratic (in square-rich region) mesophases. The disk-rich $1\Delta$ solid phase dissolves up to 30% of squares which do not have any orientational preference and are randomly distributed throughout the underlying $1\Delta$ lattice sites (Fig. S5 Sec. V). In the square-rich side, the $1\square$ solid phase dissolves up to 26% of disks and is preceded by regions of tetratic and I phases at lower pressures. For $P^* <$ 19 and for all $x_s$ values, we observed two main phase transitions: I → hexatic and hexatic → $1\Delta$ solid in the disk-rich region ($x_s < 0.3$) and I → tetratic and tetratic → $1\square$ solid in the square-rich region ($x_s > 0.75$). The transitions from the I phase to the ordered $1\Delta$ (or $1\square$) solid, occurring through an intermediate hexatic (or tetratic) mesophase are analogous to the well-studied phase transitions in the systems of pure monodisperse hard-disks (or hard-squares). The tetratic mesophase formed by the pure squares ($x_s=1.0$) is stable over a range ~ $8.25 < P^* < 15.4$ [21] that is wider than the $7.59 < P^* < 7.68$ [29] range of the hexatic mesophase formed by pure disks ($x_s=0$), a difference that can be attributed to the defects being more delocalized in the tetratic phase [21]. We found that, with increasing molar fraction of squares (disks) in the disk-rich (square-rich) region, the range of $P^*$ where the hexatic (tetratic) phase is stable increases significantly compared to the pure disk (square) system. This increase in the stability region for the hexatic phase with $x_s$ suggests that the squares accentuate the hexatic behavior as it persists even for up to $P^* \sim 16.2$, which is approximately twice the pure-disk hexatic → 1Δ solid transition pressure of $P^* \approx 7.68$. The tetratic phase is stable up to $P^* \approx 18.9$ with increasing disk concentration, which is about 1.2 times the pure-square tetratic → 1□ solid transition pressure, $P^* \approx 15.4$. This increase in the stability regions of the hexatic and tetratic phases associated with significant content of the guest component is attributable to the increased concentration of topological defects created by the dissimilarly-shaped particles residing in the host-solid lattices. These defects tend to destroy the quasi-long-range positional correlation in the solid phases in favor of the corresponding mesophase.

Figure 1 shows a peculiar continuous transition between the disk-like and the square-like behaviors over a range of square molar fractions, $0.33 < x_s < 0.6$, $14.5 < P^* < 15.7$ and $0.77 < \eta < 0.8$. We assign this region bridging the hexatic and the tetratic mesophases as the mosaic (M) phase. Along increasing $P^*$ or $\eta$, the M phase is sandwiched between the I and two-phase regions. We carried out interfacial simulations to minimize hysteretic effects and ascertain the conditions at which a single stable M phase region occurs. To characterize this phase, we analyzed the equation of state (EoS) and the six and four-fold local bond-orientational correlation functions, $g_6(r^*)$ and $g_4(r^*)$ (see Fig. 2) for $x_s = 0.5$. As can be observed in Fig. 2d, the I → M phase transition occurs at $\eta \approx 0.780$ where the global values of $\psi_6$ and $\psi_4$ increase up to 0.4-0.53, indicating significant degree of both hexatic and tetratic-like order in the system. The global $\Psi_n$ values (where $n = 4$ or 6) are evaluated by calculating the average of the $n$-fold local bond orientational order, $\Phi_n$ for all particles in the system (see SI Sec. IV for details). To distinguish the M phase from the hexatic and tetratic mesophases, we examined $g_6(r^*)$, $g_4(r^*)$ and $g(r^*)$ correlation functions (see Fig. 2). At $\eta = 0.780$, the M phase showed algebraic decay of $g_6(r^*)$ and $g_4(r^*)$ with an exponent $\approx -¼$, and short-range layering (liquid-like behavior) of $g(r^*)$. This indicates that the M phase possesses quasi-long range orientational order with both hexagonal-like and square-like structural motifs, and short-range translational order. The above results suggest that the disks and squares have comparable proclivity to form stable six-fold and four-fold ordered micro-domains, respectively, that coexist across the system. We selected the $-¼$ exponent as threshold to align with the KTHNY theory prediction for the scaling parameter lower-bound for the fluid to ($n$-fold)-atic phase transition (where $n = 4$ or 6). At $P^* = 16.5$ and $\eta = 0.80$, the $g_6(r^*)$ and $g_4(r^*)$ curves decay faster compared to the M phase; these conditions correspond to the two-phase coexistence state containing macro-segregated six and four-fold ordered domains whose bond correlation lengths are, within the given simulation box size, shorter than the M-phase quasi-long range order.

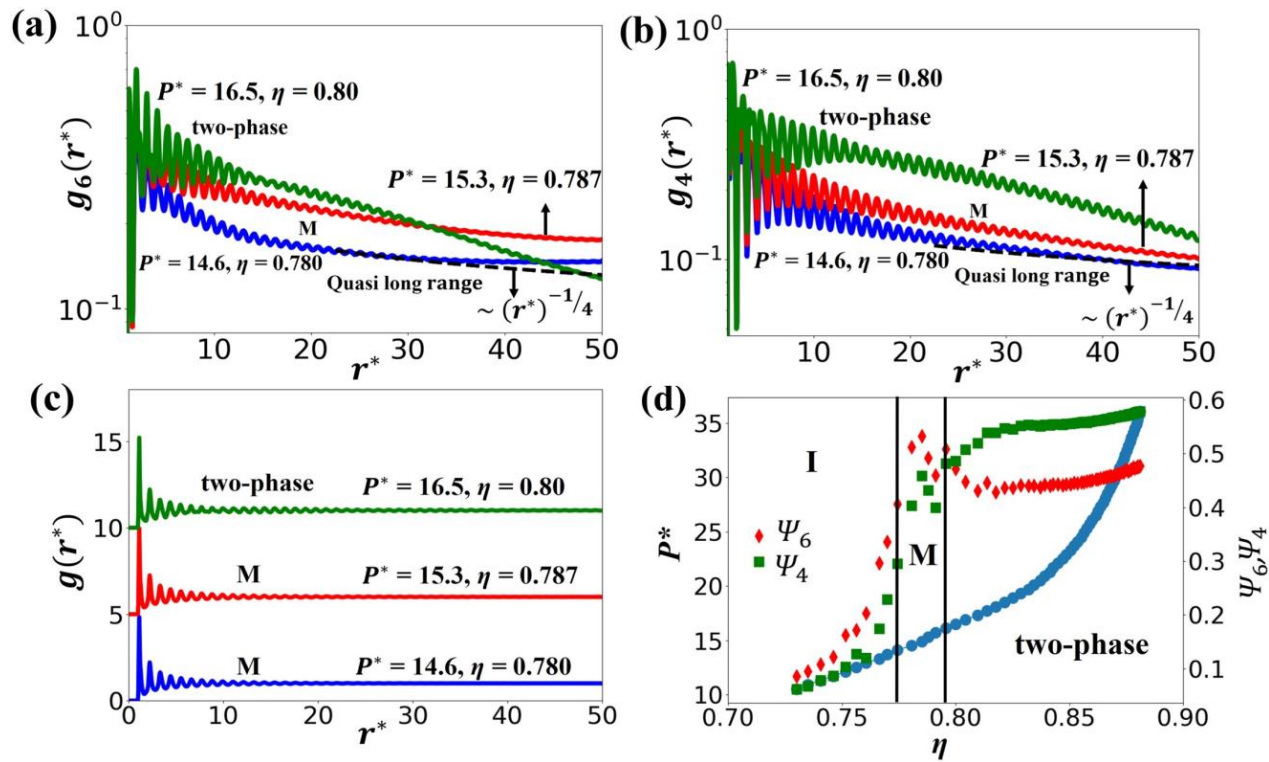

**FIG. 2** (online color). Selected properties of equimolar disk+square mixture with $\xi = 1.1$. The correlation functions (a)-(c) obtained for $N= 12,048$ particles. Bond orientational order functions $g_6(r^*)$ (a) and $g_4(r^*)$ (b) for the two-phase and M phase. The dashed line indicates algebraic decay of the orientational correlation with an exponent $\sim -\frac{1}{4}$. (c) 2D pair correlation functions shifted uniformly to distinguish peaks for the phases and conditions indicated (by pressures, $P^*$ and area fraction, $\eta$). Besides pressure $P^*$ (blue lines and circles), (d) shows $\psi_6$ (diamond) and $\psi_4$ (square) order parameters as a function of $\eta$ along with approximate phase boundaries for $N= 1600$ particles. I= isotropic phase; M= mosaic phase.

Figures 3a and 3d (inset) show configurations of the M phase and the two-phase coexistence state at $P^*= 14.9$ and $\eta= 0.783$, and $P^*= 16.5$ and $\eta = 0.80$, respectively. The clusters of six-fold and four-fold ordered domains are shown by coloring the particles based on the local values of $\Phi_6$ (Fig. 3b) and $\Phi_4$ (Fig. 3c). For the M phase, the coloring reveals a complementary correlation between the disk-rich regions with high six-fold domains and square-rich regions with high four-fold ordered domains, that are randomly distributed throughout the simulated domain. We also detected regions of RB order formed by squares with high local values of $\Phi_6$. To test that the M phase is not just a system that has become kinetically arrested *in route* to macro-phase separation, we simulated a system started at a state of complete phase separation of squares and disks at $P^*=14.9$, and confirmed that the macro-domains gradually disintegrated to form the M phase micro-domains. Movie 1 in the SI shows this transition upon decreasing $P^*$. Overall, our analysis indicates that the M phase is indeed a mesophase having a heterogeneous microstructure resembling a *"mosaic"* of different ordered micro-domains corresponding to tetratic/RB-like and hexatic-like regions.

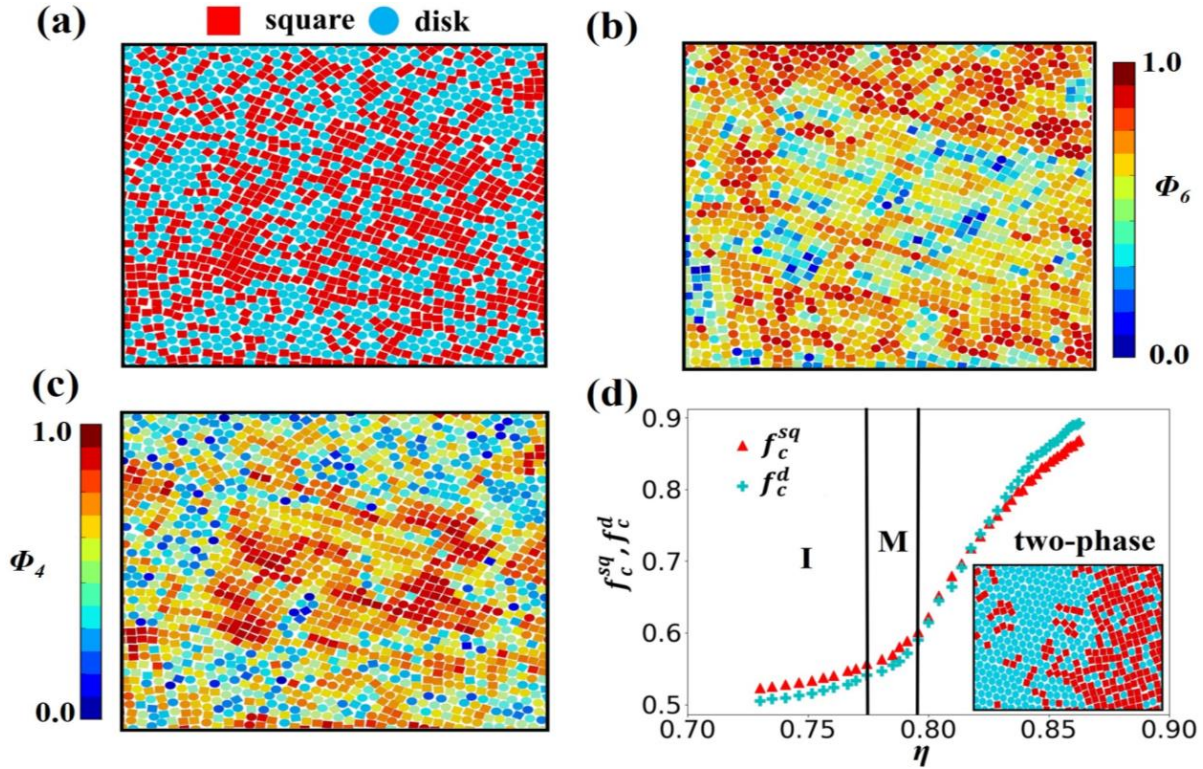

**Fig. 3** (color online). Local bond orientational and compositional order for the equimolar disks+squares mixture with $\xi = 1.1$. (a)-(c) correspond to $N= 12,048$ mosaic phase at $\eta = 0.783$ where the particles are colored based type (a) and the local values of $\Phi_6$ (b) and $\Phi_4$ (c). Each snapshot represents a section that is ~1/10$^{th}$ of the entire simulation box. (d) shows approximate phase boundaries and the local composition parameters, $f_c^{sq}$ and $f_c^d$, as a function of area fraction, $\eta$ for $N= 1600$. The inset shows a representative snapshot of two-phase coexistence state at $\eta = 0.8$. I= isotropic phase; M= mosaic phase.

To understand the mechanism associated with the I → M phase transition, we computed the local composition parameters, $f_c^{sq}$ and $f_c^d$ to detect the correlation between the local compositional heterogeneity and the presence of ordered domains formed by squares and disks (see Fig. 3d).

Parameters $f_c^{sq}$ and $f_c^d$ are the average fraction of the like-shaped nearest neighbors to a square and disk, respectively (normalized such that particles with all like-shaped neighbors corresponds to 1). For the I phase, the average values of both $f_c^{sq}$ and $f_c^d$ are close to a well-mixed value of 0.5, reflecting the overall equimolar composition. Upon compression, both $f_c^{sq}$ and $f_c^d$ increase gradually in the M phase (for $\eta > 0.77$), and then more steeply as the solid-solid phase separated region is reached ($\eta > 0.79$). The loss of the particles' local compositional mixing observed in the M phase compared to the I phase, reveals that the entropic bonding [34,35], which favors contacts between like-shaped particles, becomes sufficiently strong to seed the formation of disk-rich hexatic and square-rich tetratic/Rhombic micro-domains. The grain boundaries around these micro-domains contain particles with both $f_c^{sq}$ and $f_c^d$ values close to 0.5, which can be viewed as compositional *"defects"* contributing to the structural disorder in the M phase. The migration of these defects was monitored at $\eta = 0.783$ using *"pseudo dynamic"* Monte Carlo simulations in the *NVT* ensemble. Movie 2 in the SI shows that, although the migration of these defects is restricted to the grain boundary regions, their compositions decorrelate much faster compared to particles inside ordered domains (see Fig. S9c Sec. VI). This suggests that both the growth of ordered M domains from the I state, and the slow restructuring of the M domain patterns would be mediated by the accrual of local rearrangements at the grain boundaries.

The overall mixing entropy of the M phase, while lower than that in the I phase (where nearly ideal mixing occurs), must be significant. Indeed, while limited mixing happens at the length scale of individual particles inside clusters (as in the solid solutions) and at the grain boundaries, *'random'* mixing also occurs at the length scale of the ordered clusters. The result is a system with transient but well-defined *micro*-phase segregated regions which is quite distinct to the *macro*-phase segregated state observed at higher densities. We posit that this unique mesophase behavior engenders when, at a suitable range of compositions and densities, the two competing entropic forces, namely, entropic bonding favoring like-particle contacts and mixing entropy favoring random contacts, are in such a close balance that are able to coexist by attaining a *"compromise"* state exhibiting both segregated like-particle domains and random mixing of those domains. As the M phase is compressed to a higher density, the entropic cost of unlike contacts overpowers any gain in mixing entropy, leading to the phase separation of the individual components into disk-rich and square-rich ordered phases. Conceptually, the transitions I → M → two-solid-phases with pressure for an equimolar mixture could be seen as the coarsening in the correlation length of the ordered domains, which goes from being very short ranged (I phase), to mesoscopic (M phase) to macroscopic (two-phase state).

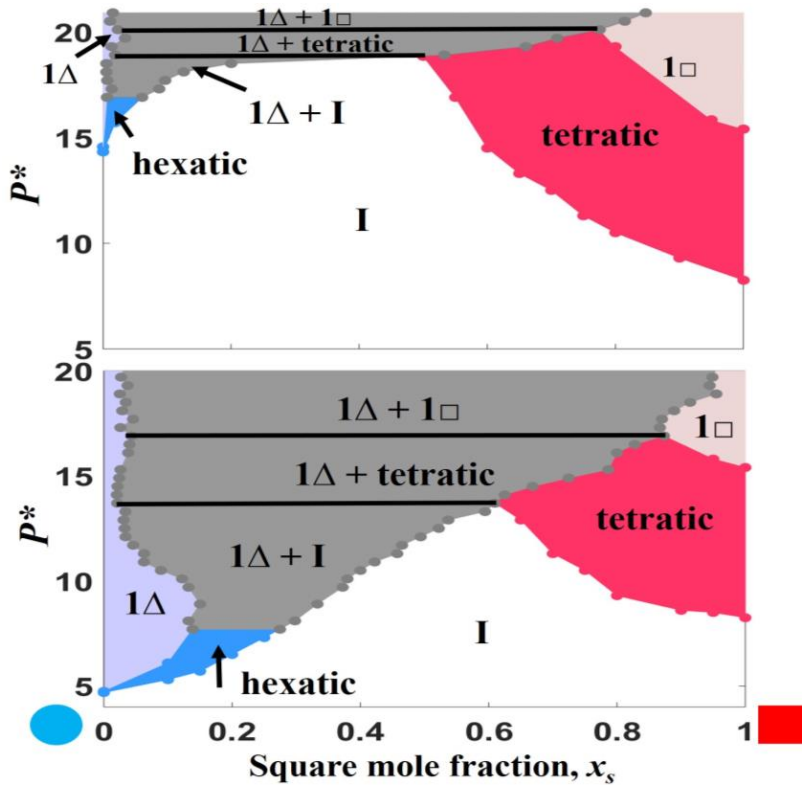

**FIG. 4** (color online). Pressure-composition phase diagrams for disks (diameter $\sigma$) and squares (side edge $a$) with different size ratios, $\xi = \sigma/a$. Top: $\xi = 0.8$, bottom: $\xi = 1.4$. $1\triangle$ = triangular solid, $1\square$ = square solid, and I = isotropic phase.

To underscore the significance of the optimal component size ratio, $\xi$, we also simulated phase diagrams for other $\xi$ values. We varied the $\xi$ values by $\pm$ 27% from the representative optimal value of 1.1 so that the associated $\Delta F_x$ values are significantly higher than those in the relatively flat region for $1.04 < \xi < 1.2$ (see Fig. S2 in SI Sec. III). Specifically, Figure 4 shows results for $\xi = 0.8$ and 1.4 for which, unlike the $\xi = 1.1$ case in Fig. 1, no M phase region was detected. In both cases, the stability region of the hexatic phase is much narrower compared to the $\xi = 1.1$ case.

Furthermore, while for the $\xi = 1.1$ case both the disk-rich and square-rich mesophases and solid solution regions are large and relatively comparable in size (giving the phase diagram a symmetric look), those regions become very asymmetric for the other $\xi$ values; i.e., the hexatic and 1Δ regions are small, especially for the $\xi = 0.8$ case. These results clearly show that a system with a (near) optimal choice of $\xi$ promotes the stability of ordered phases with substitutional disorder over wider ranges of composition and pressure and, by construction of $\Delta F_x$ [see Eq. (1)], it does so in a way that *both* pure-component ordered phases are similarly represented. Arguably, the microscopic substitutional symmetry favored by a minimal $\Delta F_x$ gets translated into a macroscopic symmetry in the substitutionally disordered solids and mesophases in the phase diagram.

While the competition between 1Δ/hexatic and 1□/tetratic ordering is not uncommon in 2D or quasi-2D systems, states resembling the M phase have only been seen under very restrictive conditions. For example, cuboctahedral nanoparticles pinned at 2D fluid-fluid interfaces have been observed to transition from a hexagonal to a square lattice only as transient, non-equilibrium states (e.g., as surface ligands are removed and particles bond through their <100> facets) [36]. 2D simulations of hard rounded squares [37] of a particular degree of roundedness have predicted the formation a *"polycrystalline"* phase with a patchy-domain structure loosely reminiscent to that of the M phase. Through the rounding of square-corners, such a system provides a physical interpolation (in a single-component system) between disks and squares to reach a state where the entropic tendencies toward the formation of hexagonal and square lattices are in close balance, like that achieved in the M phase by our disks+squares binary mixture.

For contrast, we also explored the phase behavior of a mixture of disks and hexagons with an optimal size ratio ($\xi = 1.82$, see SI sec. III) whose components have now *"compatible"* lattice symmetry as both pure-components form the hexatic and 1Δ ordered phases. The corresponding phase diagram (see Figs. S11-S12 in the SI and Sec. VIII) shows that a 1Δ solid solution and the hexatic mesophase form over the entire range of compositions.

In summary, we found a novel mosaic (M) phase bridging the disk-rich hexatic region and the square-rich tetratic region when the disk-to-square size ratio was optimized for solid-phase substitutional symmetry. It would be interesting to find out what photonic or optical properties the M phase possesses by virtue of its dual crystallinity, and whether these properties could be leveraged for applications, e.g., to fabricate a synthetic Chameleon skin [38] or optical biosensors [39]. The methods used and principles unveiled here should be general and applicable to many other mixtures.

**Acknowledgement**

Funding support from NSF award CBET-1907369 is gratefully acknowledged. The authors thank Yangyang Sun, Abhishek Sharma, Ankita Mukhtyar, and Isabela Quintela Matos for useful exchanges.